\documentclass[12pt]{article}
\usepackage{amsfonts,amssymb,amsmath}
\usepackage{theorem}
\newtheorem{definition}{Definition}[section]

\newtheorem{proposition}[definition]{Proposition}
\newtheorem{theorem}[definition]{Theorem}

\theorembodyfont{\rmfamily}

\numberwithin{equation}{section}

\def\cG{{\cal G}}                    
          \def\cK{{\cal K}}          \def\cL{{\cal L}}
                    
\def\cP{{\cal P}}                    
\def\cS{{\cal S}}          \def\cT{{\cal T}}          \def\cU{{\cal U}}
                    
\def\cY{{\cal Y}}

\newcommand{\CC}{{\mathbb C}}
\newcommand{\II}{{\mathbb I}}

\newcommand{\ZZ}{{\mathbb Z}}
\newcommand{\eps}{{\varepsilon}}

\newcommand{\finproof}{{\hfill \rule{5pt}{5pt}}}

\def\qmbox#1{{\qquad\mbox{#1}\quad}}

\def\qdet{\mathop{\rm qdet}\nolimits}
\def\sdet{\mathop{\rm sdet}\nolimits}

\newcommand{\enne}{{\cal N}}

\newcommand{\yy}{{\cal Y}}
\newcommand{\ff}{{\mathfrak f}}
\newcommand{\bs}[1]{{\boldsymbol{#1}}}
\newcommand{\wh}[1]{{\widehat{#1}}}
\newcommand{\wt}[1]{{\widetilde{#1}}}

\newcommand{\comaL}{{\cL^{*}}}
\newcommand{\comal}{L^{*}}

\begin{document}
\pagestyle{empty}
\renewcommand{\thefootnote}{\fnsymbol{footnote}}
\setcounter{footnote}{1}

%%%%%%%%%%%%%%%%%%%%%%%%%%%%%%%%%
%%%%%  HEADINGS POUR DRAFT  %%%%%Universal treatment of 
%%%%%%%%%%%%%%%%%%%%%%%%%%%%%%%%%
% \markright{\today\dotfill DRAFT\dotfill }
% \pagestyle{myheadings}

\begin{center}

{\Large \textsf{Analytical Bethe Ansatz for open spin chains\\[1.2ex]
   with soliton non preserving boundary conditions}}

\vspace{10mm}

{\large D. Arnaudon\footnote{arnaudon@lapp.in2p3.fr,
nc501@york.ac.uk, doikou@lapp.in2p3.fr, frappat@lapp.in2p3.fr,
ragoucy@lapp.in2p3.fr}$^{\,a}$, N. Cramp{\'e}$^{\dag\, b}$,
A. Doikou$^{\dag \,a}$, \\[1.2ex]
L. Frappat$^{\dag\, a,c}$ and
{\'E}. Ragoucy$^{\dag\, a}$}

\vspace{10mm}

\emph{$^{a}$ Laboratoire d'Annecy-le-Vieux de Physique Th{\'e}orique}

\emph{LAPTH, CNRS, UMR 5108, Universit{\'e} de Savoie}

\emph{B.P. 110, F-74941 Annecy-le-Vieux Cedex, France}

\vspace{7mm}

\emph{$^b$ University of York, Department of mathematics}

\emph{Heslington, York YO10 5DD, United Kingdom}

\vspace{7mm}

\emph{$^c$ Member of Institut Universitaire de France}

\end{center}

\vfill

\begin{abstract}
We present an ``algebraic treatment'' of the analytical Bethe ansatz for open spin chains with soliton non preserving (SNP) boundary conditions. For this purpose, we introduce abstract monodromy and transfer 
matrices which provide an algebraic framework for the analytical 
Bethe ansatz. It allows us to deal with a generic $gl(\enne)$ open SNP
spin chain possessing on each site an arbitrary representation.
As a result, we obtain the Bethe equations in their full generality. The classification of finite 
dimensional irreducible representations for the twisted Yangians are directly 
linked to the calculation of the transfer matrix eigenvalues. 
\end{abstract}

\vfill
\begin{center}
\end{center}
\vfill

\rightline{\texttt{math-ph/0503014}}
\rightline{LAPTH-1090/05} 
\rightline{March 2005}

\baselineskip=16pt

%%%%%%%%%%%%%%%%%%%%%%%%%%%%%%%%%%%%%%%%%%%%%%%%%%%%%%%%%%%%%%%%%%%%%%%%%%%%%%%
\newpage
\pagestyle{plain}
\renewcommand{\thefootnote}{\arabic{footnote}}
\setcounter{footnote}{0}
\setcounter{page}{1}

\section*{Introduction}

A great interest for generalizations of integrable quantum spin
chains has recently appeared. This renewed interest is primarily due to  new applications 
of these spin chains in various fields, such as 
condensed matter \cite{gv,jjgj,br,pbas,wbf,kmyu,jlm,gut}, integrable relativistic quantum field theories
\cite{mw,aff,zaza,fsz}, quantum chromodynamics theory \cite{lip, FaKo} 
or AdS/CFT correspondence
\cite{miza,beis,Zarembo}. In this context, an
important development was the definition of integrable quantum spin
chain with non-trivial boundaries. Usually, in the framework of 
spin chain models, these boundaries are characterized by a matricial
solution to the reflection equation \cite{cherednik,sklyanin}.
However, ``new" types of boundaries have been introduced in
\cite{doikou1,aacdfr, doikou1b} where the matrix is now solution of the
so-called soliton non preserving (SNP) reflection equation. This equation was
intensively studied from a mathematical point of view to define the
algebras called twisted Yangians \cite{Ytwist,MNO}. Later, it has been
used to introduce boundaries in the affine Toda field
theories \cite{boco,gand,dede,dema}.

In our previous work \cite{byebye}, we used the classification of
irreducible finite dimensional
representations of the Yangian and of the reflection algebra to
construct and solve spin chains where each sites are associated to a
different representation of $gl(\enne)$. In the present paper, we
extend this procedure to the SNP case. Indeed, the classification of the
representations of the twisted Yangian provided in \cite{twmolev}
allows us to obtain the spin chain with non-soliton preserving
boundaries where each spin of the chain can be in different representations.

This article is organized as follows. In the first three sections we
 recall elementary notions on the Yangian of $gl(\enne)$,
the twisted Yangian and the representations of the Yangian. These
sections provide 
 all the definitions needed for the article to be
self-contained. We also introduce the transfer matrix and 
determine its symmetry algebra, which is consequently the one of the 
considered spin chain. In 
section \ref{spinSNP}, the classification of the twisted Yangian are
described, as well as the fusion procedure, to obtain constraints on
the transfer matrix. Using these constraints, we then determine the
dressing functions for a general transfer matrix eigenvalue. Finally, 
we compute the Bethe
equations, by analytical Bethe ansatz, for the general SNP $gl(\enne)$
spin chain.

\section{Yangian $\yy(gl(\enne)$  \label{sect:algebra}}

We will consider the $gl(\enne)$ invariant $R$ matrices
\cite{yang,baxter}
\begin{eqnarray}
  R_{ab}(\lambda) =\II_\enne \otimes \II_\enne -
\frac{\hbar\;\cP_{ab}}{\lambda} \;, \label{r}
\end{eqnarray}
where $\cP_{ab}$ is the permutation operator
\begin{equation}
  \label{eq:P12}
  \cP_{ab} = \sum_{i,j=1}^\enne E_{ij} \otimes E_{ji}
\end{equation}
and $\hbar$ is the free deformation parameter. It satisfies the following
properties \\[2mm]
\textit{(i) Yang--Baxter equation}
\cite{mac,yang,baxter,korepin,KoIzBo}
\begin{eqnarray}
  R_{ab}(\lambda_{a}-\lambda_{b})\ R_{ac}(\lambda_{a})\ R_{bc}(\lambda_{b})
  =R_{bc}(\lambda_{b})\ R_{ac}(\lambda_{a})\
  R_{ab}(\lambda_{a}-\lambda_{b})
  \label{YBE}
\end{eqnarray}
\textit{(ii) Unitarity}
\begin{eqnarray}
  R_{ab}(\lambda)\ R_{ba}(-\lambda) = \zeta(\lambda)\,
  \II_\enne \otimes \II_\enne\label{uni1}\;,
\end{eqnarray}
where $R_{ba}(\lambda) =\cP_{ab} R_{ab}(\lambda) \cP_{ab} =
R_{ab}^{t_{a}t_{b}}(\lambda) = R_{ab}(\lambda)$ and
\begin{eqnarray}
  \zeta(\lambda) = \left(1-\frac{\hbar}{\lambda}\right)
\left(1+\frac{\hbar}{\lambda}\right)\;.
\end{eqnarray}
It obeys $[A_{a} A_{b},\ R_{ab}(\lambda)] = 0$ for $A\in End(\CC^\enne)$. \\

The Yangian $\yy(gl(\enne))$ \cite{Drinfeld} is the complex associative unital
algebra with the generators $\{L_{ij}^{(n)}|1\leq i,j\leq \enne,
n\in \ZZ_{\geq 0}\}$
 subject to the defining relations
\begin{eqnarray}
\label{relcom}
[L_{ij}^{(r+1)}\,,\,L_{kl}^{(s)}]-[L_{ij}^{(r)}\,,\,L_{kl}^{(s+1)}]
=L_{kj}^{(r)}\,L_{il}^{(s)}-L_{kj}^{(s)}\,L_{il}^{(r)}\;,
\end{eqnarray}
where $r,s\in \ZZ_{\geq 0}$ and $L_{ij}^{(0)}=\delta_{ij}$. These
relations are encoded in a simple equation, called FRT exchange
relation \cite{FRT}
\begin{eqnarray}
\label{RTT}
R_{ab}(\lambda_a-\lambda_b)\;\cL_a(\lambda_a)\;\cL_b(\lambda_b)=
\cL_b(\lambda_b)\;\cL_a(\lambda_a)\;R_{ab}(\lambda_a-\lambda_b)\;,
\end{eqnarray}
where the generators are gathered in the following matrix (belonging
to $End(\CC^\enne)\otimes \yy(gl(\enne))[[\lambda^{-1}]]$)
\begin{eqnarray}
\label{def:T} \cL(\lambda)=\sum_{i,j=1}^\enne E_{ij}\otimes
L_{ij}(\lambda) =\sum_{i,j=1}^\enne E_{ij}\otimes \sum_{r \geq 0}
\frac{\hbar^{r}}{\lambda^r}~L_{ij}^{(r)}= \sum_{r \geq 0}
\frac{\hbar^{r}}{\lambda^r}~\cL^{(r)} \;.
\end{eqnarray}

The quantum determinant $\qdet\cL(\lambda)$ is a formal series in
$\lambda^{-1}$ with coefficients in $\yy(gl(\enne))$ defined as
follows
\begin{eqnarray}
    \label{qdet}
    \qdet\cL(\lambda)= \sum_{\sigma\in \mathfrak{S}_\enne}sgn(\sigma)~
    L_{1,\sigma(1)}(\lambda-\hbar\enne+\hbar) \cdots
    L_{\enne,\sigma(\enne)}(\lambda)\;,
\end{eqnarray}
where $\mathfrak{S}_\enne$ is the permutation group of $\enne$
indices. A well-known result (see e.g. \cite{molev}) establishes
that the coefficients of $\qdet\cL(\lambda)$ are algebraically
independent and generate the center of $\yy(gl(\enne))$. There
exists an equivalent definition of the quantum determinant which
will be used in the following as well:
\begin{equation}
\label{qdet2}
\qdet\cL(\lambda)\;A_\enne=\cL_\enne(\lambda-\hbar\enne+\hbar)
\cdots \cL_1(\lambda)\;A_\enne\;.
\end{equation}
where $A_\enne$ is the antisymmetriser operator, a one-dimensional
projector in $(\CC^\enne)^{\otimes \enne}$, i.e.
\begin{equation}
\label{antisym} A_m(e_{1}\otimes\cdots\otimes
e_{m})=\frac{1}{m!} \sum_{\sigma\in\mathfrak{S}_m}
sgn(\sigma)~e_{{\sigma(1)}}\otimes\cdots\otimes
e_{{\sigma(m)}}\;.
\end{equation}

For the study of spin chains, it is useful to introduce the following morphisms
of $\yy(gl(\enne))$:
\begin{description}
\item[Inversion]
\begin{eqnarray}
inv:\ \cL(\lambda)\longmapsto \cL^{-1}(\lambda)
\end{eqnarray}
\item[Sign]
\begin{eqnarray}
 sg:\ \cL(\lambda)\longmapsto \cL(-\lambda)\,,\label{sign}
\end{eqnarray}
\item[Transposition]
\begin{eqnarray}
 t:\ \cL_a(\lambda)\longmapsto \cL_a^{t_a}(\lambda)\,,\label{transp}
\end{eqnarray}
where $^{t_a}$ is a generalised transposition (defined below) acting in space $a$ only.
\item[Shift]
\begin{eqnarray}
 s_{a}:\ \cL(\lambda)\longmapsto \cL(\lambda+a)\,,\ a\in\CC\label{shift}
\end{eqnarray}
\end{description}
The three first mapping are idempotent algebra anti-morphisms, while the last
one is an algebra automorphism.\\
The
generalised transposition $^t$, which depends on a sign $\theta=\pm1$, is
related to the usual transposition $^T$ by, for any matrix $A$,
\begin{equation}
A^t=V^{-1}\,A^T\,V \;\;\mbox{where}\left\{
\begin{array}{ll}
  V = \mbox{antidiag}(1,1,\ldots,1)\,, &\ \mbox{ for which }\
  V^2=\theta=1\\
  \mbox{or}& \\
  V=\mbox{antidiag}\Big(
  \underbrace{1,\ldots,1}_{\enne/2}\,,\,\underbrace{-1,\ldots,-1}_{\enne/2}
    \Big)\,,
    &\mbox{ for which }\
  V^2=\theta=-1\,.
\end{array}\right.
\label{eq:V}
\end{equation}
The second case is forbidden when $\enne$ is odd.
\\
The elements of $\cL^{-1}(\lambda)$ are computed in terms of $\cL(\lambda)$
by
\begin{eqnarray}
    \label{comp:inv}
    \cL^{-1}(\lambda-\hbar\enne+\hbar)=\big(\qdet\cL(\lambda)\big)^{-1}\;
    \comaL(\lambda)\;,
\end{eqnarray}
where $\comaL(\lambda)$ is the quantum comatrix, i.e. the entries
$\comal_{ij}(\lambda)$ of $\comaL(\lambda)$ are $(-1)^{i+j}$ times the
quantum determinants of the submatrices of $\cL(\lambda)$ obtained by
removing the $i^{th}$ column and $j^{th}$ row. \\

In the following, $\cL(\lambda)$ will be denoted by $\cL_{ap}(\lambda)$
since we deal with tensor products of Yangian. The index $a$ is the
auxiliary space index (i.e. the space on which $E_{ij}$ acts) and $p$ is
the quantum space index (i.e. the space on which $L_{ij}^{(r)}$ acts).
Starting from local operators $\cL_{ap}(\lambda)$ ($1 \le p \le \ell$)
acting on different spaces, one constructs a non-local algebraic object,
the monodromy matrix
\begin{align}
\label{mono} \cT_a(\lambda)&=
\cL_{a1}(\lambda)\;\cL_{a2}(\lambda)\;\dots\;\cL_{a\ell}(\lambda)\in
End (\CC^\enne) \otimes \left(\yy(gl(\enne))\right)^{\otimes \ell}
\end{align}
which satisfies the defining relations of the Yangian
\begin{eqnarray}
\label{Rtt}
R_{ab}(\lambda_a-\lambda_b)\;\cT_{a}(\lambda_a)\;\cT_{b}(\lambda_b)=
\cT_{b}(\lambda_b)\;\cT_{a}(\lambda_a)\;R_{ab}(\lambda_a-\lambda_b)\;.
\end{eqnarray}

\section{Algebraic transfer matrix for twisted Yangian}

\subsection{${K}$ matrix}

In order to study open spin chains with soliton non-preserving boundary
conditions, we need to use subalgebras of Yangians called twisted Yangians,
$\cY^{\pm}(\enne)$ \cite{MNO}. The sign $+$ or $-$ allows us to choose
between the two types of twisted Yangians (orthogonal or symplectic).
First, we need to introduce numerical matrices, called ${K}$ matrices,
which are solutions to the soliton non-preserving reflection
equation\footnote{There exist different definitions of this relation
depending upon the shift $\rho$ of the spectral parameter in
$R^{t_a}_{ab}$\label{ft:shift}.} \cite{MNO}
\begin{eqnarray}
&&R_{ab}(\lambda_{a}-\lambda_{b})\ {K}_{a}(\lambda_{a})\
  R_{ab}^{t_a}(-\lambda_{a}-\lambda_{b}-\hbar\rho)\ {K}_{b}(\lambda_{b})=
  \nonumber\\
  &&\hspace{3cm}
  {K}_{b}(\lambda_{b})\
  R_{ab}^{t_a}(-\lambda_{a}-\lambda_{b}-\hbar\rho)\ {K}_{a}(\lambda_{a})\
  R_{ba}(\lambda_{a}-\lambda_{b}) \;,\qquad
  \label{retw}
\end{eqnarray}
In this case, the ${K}$ matrix is interpreted as the reflection of a
soliton on the boundary, coming back as an anti-soliton. The solutions of
(\ref{retw}) have been classified in \cite{aacdfr}:
\begin{proposition}
\label{prop:SNP1} Any invertible solution of the soliton non-preserving
reflection equation (\ref{retw}) is a constant matrix (up to a
multiplication by a scalar function) such that $K^t = \varepsilon K$ with
$\varepsilon=\pm 1$.
\end{proposition}
In the following, we restrict ourselves to the case where ${K}_a$ is
diagonal and is given by
\begin{eqnarray}
    \label{eq:Kmatrix}
    {K}=diag(\zeta_1,\dots,\zeta_\enne)\qmbox{with}
    \zeta_{\overline{k}}=\varepsilon~\zeta_k\,,\quad \zeta_k\neq
    0\qmbox{and} \eps=\pm1\,.
\end{eqnarray}
The variables $\zeta_{k}$ are $\left[\frac{\enne+1}{2}\right]$ free
parameters of the model.
\\
Let us remark that when $\enne$ is odd, one must take $\eps=+1$ to ensure
the invertibility of $K$ (and $\theta=+1$ in this case), while
for $\enne$ even, there are four choices ($\theta=\pm1$, $\eps=\pm1$).

\subsection{Twisted Yangians}

The twisted Yangians are constructed as subalgebras of the Yangian
$\yy(gl(\enne))$. Starting from the generators $\cT(\lambda)$ of
$\yy(gl(\enne))$ introduced in (\ref{def:T}), we define
\begin{equation}
\label{S-T}
  \cS_a(\lambda)=\cT_a(\lambda)\,{K}_a
\,{\cT_a}^{t_a}(-\lambda-\hbar\rho)\;.
\end{equation}
$\cS(\lambda)$ generates the algebra $\yy^{\,\theta\eps}(\enne)$
whose exchange relations$^{\ref{ft:shift}}$ are given by
\begin{eqnarray}
&& R_{ab}(\lambda_{a}-\lambda_{b})\ \cS_{a}(\lambda_{a})\
  R^{t_a}_{ab}(-\lambda_{a}-\lambda_{b}-\hbar\rho)\ \cS_{b}(\lambda_{b})=
  \nonumber\\
  &&\hspace{3cm}\cS_{b}(\lambda_{b})\
  R^{t_a}_{ab}(-\lambda_{a}-\lambda_{b}-\hbar\rho)\ \cS_{a}(\lambda_{a})\
  R_{ba}(\lambda_{a}-\lambda_{b}) \;.\qquad
  \label{retw-algebra}
 \end{eqnarray}
This relation is a direct consequence of (\ref{shift}), (\ref{transp}) and
(\ref{retw}). The matrix $\cS(\lambda)$ satisfies a supplementary symmetry
relation
\begin{eqnarray}
\label{sym-rel}
\cS^{t_a}_a(\lambda)=\varepsilon~\cS_a(-\lambda-\hbar\rho)-
\frac{\theta\;\hbar}{2\lambda+\hbar\rho}~
\big(\cS_a(-\lambda-\hbar\rho)-\cS_a(\lambda)\big)\;.
\end{eqnarray}
As in equation (\ref{def:T}), we define
\begin{equation}
\cS(\lambda)=\sum_{i,j=1}^\enne E_{ij}\otimes S_{ij}(\lambda)
=\sum_{n=0}^{+\infty} \frac{\cS^{(n)}}{\lambda^{n}}~.
\end{equation}
The commutation relations (\ref{retw-algebra}) and the symmetry
relation (\ref{sym-rel}) show that $\cS^{(1)}$ generates a
$so(\enne)$ (resp. a $sp(\enne)$) algebra when $\theta\varepsilon=1$
(resp. $\theta\varepsilon=-1$),
subalgebras in $\yy^{\,\theta\eps}(\enne)$ \cite{MNO}.  \\

One can show that $\yy^{\,\theta\eps}(\enne)$ has a non-trivial center
generated by the coefficients of the following series, the so-called
Sklyanin determinant $\sdet\cS(\lambda)$ defined by
\begin{eqnarray}
    \cS_{<a_{\enne}\ldots a_{1}>}(\lambda)\,A_{\enne}=
    A_{\enne}\,\cS_{<a_{\enne}\ldots a_{1}>}(\lambda)\,A_{\enne}=
    \sdet\cS(\lambda)\,A_{\enne}
    \label{sdetSNP}
\end{eqnarray}
where
\begin{equation}
    \cS_{<a_{\enne}\ldots a_{1}>}(\lambda)= \left(\prod_{2\leq k\leq
    \enne}^{\longleftarrow} \cS_{a_{k}}(\lambda_{k})\,
    R^{t_{a_{k}}}_{a_{k}a_{k-1}}(-\lambda_{k}-\lambda_{k-1}-\hbar\rho)\ldots
    R^{t_{a_{k}}}_{a_{k}a_{1}}(-\lambda_{k}-\lambda_{1}-\hbar\rho)\right)
    \, \cS_{a_{1}}(\lambda_{1})
    \label{SfusansK}
\end{equation}
and $\lambda_{k}=\lambda-\hbar(k-1)$. It satisfies the following relation
\begin{eqnarray}
    \sdet\cS(\lambda) = \sdet{K}(\lambda) ~~ \qdet\cT(\lambda) ~~
    \qdet\cT(-\lambda-\hbar(\rho-\enne+1))\;,
    \label{sdet-qdet}
\end{eqnarray}
with
\begin{eqnarray}
\sdet{K}(\lambda)= \zeta_1\zeta_2\dots\zeta_\enne \;
\frac{2\lambda-(\theta\varepsilon+1)n\hbar+\hbar+\hbar\rho}
{2\lambda-2n\hbar+\hbar+\hbar\rho} \qmbox{and} \displaystyle
n=\left[\frac{\enne}{2}\right]\,.
\end{eqnarray}

\subsection{Transfer matrix and symmetry of the model}

The transfer matrix is defined by
\begin{eqnarray}
s(\lambda)=tr_a\left(\cS_a(\lambda)\right)= \sum_{i=1}^\enne
S_{ii}(\lambda)
\end{eqnarray}
and satisfies a crossing relation deduced from (\ref{sym-rel})
\begin{eqnarray}
\label{cross} s(\lambda)=
\frac{2\lambda\;\varepsilon+\hbar(\rho\;\varepsilon-\theta)}
{2\lambda+\hbar(\rho-\theta)} ~~s(-\lambda-\hbar\rho)\;.
\end{eqnarray}
This relation differs from the one given in \cite{aacdfr}, where
the case $\varepsilon=+1$ is treated in the fundamental
representation, because of the definitions (\ref{S-T}) and
(\ref{retw-algebra}) used here (see footnote \ref{ft:shift}). The
commutation relations defining the twisted Yangian allows us to
show
\begin{eqnarray}
\label{comtmyt} [s(\lambda),s(\mu)]=0\;.
\end{eqnarray}
This commutation of the transfer matrix guarantees the integrability
of the model characterized by the Hamiltonian which is any linear
combination of the coefficients of $s(\lambda)$. The following
proposition gives the symmetry of the model.

\begin{proposition}
    \label{prop:22}
    For a generic $K$ matrix, the transfer matrix $s(\lambda)$ describing
    soliton non-preserving open spin chain models admits as symmetry
    algebra the direct sum of $\left[\frac{\enne+1}{2}\right]$ copies of
    $\cG$ where
    \begin{equation}
        \cG =
        \begin{cases}
            sp(2) \qmbox{algebras when} \theta=-1\mbox{ and
            }\varepsilon=1\\
            so(2) \qmbox{algebras when} \theta=1\mbox{ and }\varepsilon=1\\
            U(1) \qmbox{algebras when $\eps=-1$.}
        \end{cases}
    \end{equation}
    If $K$ contains $p$ equal parameters in the set $\{\zeta_{j} \;\vert\;
    1 \le j \le \enne/2\}$, then the corresponding symmetry subalgebra
    generated by the direct sum of the $p$ algebras $sp(2)$ (resp. $so(2)$
    or $U(1)$) is enlarged to an $sp(2p)$ (resp. $so(2p)$ or $U(p)$)
    algebra. \\
    In the particular case where $\enne=2n+1$ and $\zeta_{n+1}$ is equal to
    the above $p$ parameters, the symmetry subalgebra generated by the
    direct sum of the $p+1$ algebras $so(2)$ (resp. $U(1)$) is enlarged to
    an $so(2p+1)$ (resp. $U(p+1)$) algebra.
\end{proposition}
\textbf{Proof:} Starting from (\ref{retw-algebra}), taking the trace in
space $a$ and looking at the coefficient of $\lambda_{b}^{-1}$ and
$E_{ij}$, one gets
\begin{equation}
    {\left[\,s(\lambda)\,,\,S^{(1)}_{ij}\,\right]} =
\hbar\,(\zeta_{i}-\zeta_{j})\,\Big(S_{ij}(\lambda)
+(\cS^t)_{ij}(\lambda)\Big)\,.
\end{equation}
This shows that when $\zeta_{i}=\zeta_{j}$, $S^{(1)}_{ij}$ commutes
with $s(\lambda)$, which leads to the different cases given in the
proposition, taking into account the algebra generated by
$\cS^{(1)}$ (see above).
\finproof\\
Let us remark that in proposition \ref{prop:22} all the symmetry algebras
have the same rank $\left[\frac{\enne+1}{2}\right]$.

\section{Representations \label{reps}}

In the following, it will be necessary to find the irreducible
finite-dimensional representations of the
twisted Yangians in order to construct soliton non-preserving open spin
chains. For such a purpose, we reproduce the techniques used in \cite{twmolev}
to classify these representations.
The first step consists in constructing the representations of the
local operators $\cL_{ap}$. We then deal with the tensor product of these
representations using (\ref{mono}). Finally, we obtain representation of
the twisted Yangian by (\ref{S-T}). 

\subsection{Evaluation representation of local operator}

Let $\{e_{ij}\}$ be a basis of the Lie algebra $gl(\enne)$. The
finite-dimensional irreducible representation of $gl(\enne)$, $M({\bs
\alpha})$, with highest weight ${\bs \alpha} =
(\alpha_1,\dots,\alpha_\enne)$ and associated to the highest weight vector
$v$ is characterized by
\begin{eqnarray}
    &&e_{kj}\;v=0 \qmbox{,} 1\leq k<j \leq \enne  \\
    &&e_{kk}\;v=\alpha_k\; v \qmbox{,}  1\leq k \leq \enne \;,
\end{eqnarray}
where $\alpha_1,\dots,\alpha_\enne\in \CC$. \\
The following algebra homomorphism from $\yy(gl(\enne))$ to
$\cU(gl(\enne))$ (universal enveloping algebra of $gl(\enne)$) \footnote{To
be compatible with the pseudo-vacuum as usually defined in the study of
spin chain models, the convention used here for the homomorphism differs
from the one introduced in \cite{twmolev}. The link between the two
conventions is provided by the Yangian automorphism $T(\lambda)\longmapsto
T^T(-\lambda)$.}
\begin{eqnarray}
    \label{eval}
    L_{ij}(\lambda)\longmapsto
    \delta_{ij}-\frac{\hbar\;e_{ji}}{\lambda}\;,
\end{eqnarray}
allows us to build the evaluation representation $M_{\lambda+a}({\bs
\alpha})$ of $\yy(gl(\enne))$ from $M({\bs \alpha})$ satisfying
\begin{eqnarray}
    &&L_{jk}(\lambda)\;v=0 \qmbox{,} 1\leq k<j \leq \enne  \\
    &&L_{kk}(\lambda)\;v=\left(1-\frac{\hbar\;\alpha_k}{\lambda+a}\right)\;
    v \qmbox{,} 1\leq k \leq \enne \;.
\end{eqnarray}
It is important for the following to remark that the previous relations
imply that the entries of the matrix $(\lambda+a) \cL(\lambda)$ are
analytical. \\
The representation $M_\lambda((1,0,\dots,0))$ associated to the $gl(\enne)$
fundamental representation of $\cL(\lambda)$ provides the $R$ matrix
(\ref{r}).

\subsection{Representation of the Yangian}

The evaluation representations of $\cL(\lambda)$ allow us to build a
representation of $\cT(\lambda)$. Indeed, evaluating each of the local
operator $\cL_{a n}(\lambda)$ in a representation
$M_{\lambda+a_{n}}(\bs{\alpha^n})$ for $1 \le n \le \ell$, the tensor
product built on
\begin{eqnarray}
    \label{tensorp}
    M_{\lambda+a_{1}}(\bs{\alpha^1}) \otimes \dots \otimes
    M_{\lambda+a_{\ell}}(\bs{\alpha^\ell})
\end{eqnarray}
provides a finite-dimensional representation for $\cT(\lambda)$.\\
Denoting by $v^n$ the highest weight vector associated to
$\bs{\alpha^n}=(\alpha^n_1,\dots,\alpha^n_\enne)$, the vector
\begin{equation}
    v^+=v^1 \otimes \dots \otimes v^\ell
    \label{v+}
\end{equation}
is the highest weight vector of the representation (\ref{tensorp}), i.e.
\begin{eqnarray}
    &&T_{jk}(\lambda)\;v^+=0 \qmbox{,} 1\leq k<j \leq \enne \\
    &&T_{kk}(\lambda)\;v^+=\prod_{n=1}^\ell
    \left(1-\frac{\hbar\;\alpha^n_k}{\lambda+a_{n}}\right)\;v^+ \qmbox{,}
    1\leq k \leq \enne \;.
    \label{HWmono2}
\end{eqnarray}
For later convenience, we introduce the following polynomials, so-called
Drinfeld polynomials:
\begin{eqnarray}
    \label{drinP} P_k(\lambda)=\prod_{n=1}^\ell
    \left(\lambda+a_{n}-\hbar\;\alpha^n_k\right)\;.
\end{eqnarray}
We will be interested only in the irreducible finite-dimensional
representations of the monodromy matrix. Indeed, when the representation is
reducible, the Bethe ansatz does not give all the eigenvalues of the
transfer matrix. There exists a necessary and sufficient criterion for a
tensor product of Yangian representations to be irreducible
\cite{molirr,byebye}.

\section{Spin chains with non preserving boundary conditions \label{spinSNP}}

\subsection{Representations of $\yy^{\,\theta\eps}(\enne)$}

The monodromy matrix used to construct open spin chains with non-preserving
boundary conditions is constructed from the realization of the Yangian
(\ref{mono}) and takes the following form
\begin{eqnarray}
\label{monotw} \cS_a(\lambda)=
\cL_{a1}(\lambda)\;\dots\;\cL_{a\ell}(\lambda)~ {K}_a~
\cL_{a\ell}^{t_a}(-\lambda-\hbar\rho)\;\dots\;
\cL_{a1}^{t_a}(-\lambda-\hbar\rho)\;.
\end{eqnarray}
In order to study the representations of $\yy^{\,\theta\eps}(\enne)$, we
start from the Yangian representations introduced in section \ref{reps}.
Let $M_\lambda({\bs \alpha})$ be an evaluation representation of
$\cL(\lambda)$ with the highest weight vector $v$. We easily see that $v$
is the highest weight vector of $\cL^{t}(\lambda)$ with
\begin{eqnarray}
\label{HWtran1}
&&\overline{L}_{jk}(\lambda)\;v
={L}_{\overline{kj}}(\lambda)\;v
=0 \,,\qquad 1\leq k<j \leq \enne  \\
\label{HWtran2}
&&\overline{L}_{kk}(\lambda)\;v=
{L}_{\overline{kk}}(\lambda)\;v=
\left(1-\frac{\hbar\;\alpha_{\overline{k}}}{\lambda}
\right)
\; v
\,,\qquad   1\leq k \leq \enne \;,\quad
\end{eqnarray}
where $\overline{L}_{jk}(\lambda)$ are the matrix elements of the matrix
$\cL^{t}(\lambda)$ and the overlined indices $1 \le k \le \enne$ are
defined by
\begin{eqnarray}
    \overline{k}=\enne+1-k\;.
\end{eqnarray}
It is known \cite{twmolev} that any finite-dimensional representation of
$\yy^{\,\theta\eps}(\enne)$ is a highest weight representation. They can be
constructed in the following way:

\begin{theorem}
Let us consider the Yangian highest weight representation
    $M_{\lambda+a_1}(\bs{\alpha^1})\otimes \ldots\otimes
    M_{\lambda+a_\ell}(\bs{\alpha^\ell})$ with highest weight vector
    $v^+=v^1\otimes\ldots\otimes v^\ell$. We change the normalization in
    order to have analytical eigenvalues:
    \begin{equation}
        \wh\cS_a(\lambda) =
        \prod_{i=1}^\ell(\lambda+a_i)(-\lambda-\hbar\rho+a_i)~\cS_a(\lambda)
        \qmbox{and} \wh s(\lambda)=tr_{a}\wh\cS_{a}(\lambda)\,.
    \end{equation}
    Then the realization (\ref{monotw}) generates a
    $\yy^{\,\theta\eps}(\enne)$ highest weight representation, whose highest
    weight vector is also $v^+$ with
    \begin{eqnarray}
        \label{HWmonotran1}
        &&\wh S_{jk}(\lambda)\;v^+ = 0 \,,\qquad 1\leq k<j \leq \enne \\
        \label{HWmonotran2}
        &&\wh S_{kk}(\lambda)\;v^+ = \zeta_k \sigma_k(\lambda)\; v^+
        \,,\qquad 1\leq k \leq \frac{\enne+1}{2} \;, \\
        && \label{HWmonotran3} \wh S_{kk}(\lambda)\;v^+=
        \frac{\zeta_k}{2\lambda+\hbar\rho} \Big(
        \big(2\lambda+\hbar(\rho-\theta\varepsilon)\big) \sigma_k(\lambda)
        +\theta\varepsilon\hbar~\sigma_{\overline{k}}(\lambda)\Big) \; v^+
        \,, \qquad \frac{\enne}{2}+1\leq k \leq \enne
    \end{eqnarray}
    where
    \begin{eqnarray}
        \label{sigma}
        \sigma_k(\lambda)=P_k(\lambda)P_{\overline{k}}(-\lambda-\hbar\rho)\;.
    \end{eqnarray}
    Let us recall that the $\zeta_{i}$ are the elements of the diagonal $K$ matrix satisfying eq. (\ref{eq:Kmatrix}).
\end{theorem}
\textbf{Proof:} The equalities (\ref{HWmonotran1}) and
(\ref{HWmonotran2}) are a consequence of (\ref{HWtran1}) and
(\ref{HWtran2}) as well as of a direct computation using the Yangian
commutation relations. The last relation (\ref{HWmonotran3}) is
computed from the previous ones using (\ref{sym-rel}). \finproof\\
Remark that the functions $\sigma_{k}(\lambda)$ obey the crossing
relations
\begin{equation}
        \sigma_{\bar k}(-\lambda-\hbar\rho)=\sigma_{k}(\lambda)\,.
        \label{crossig}
\end{equation}
For a given representation, the Sklyanin determinant is given by
\begin{eqnarray}
\sdet\wh\cS(\lambda) = \sdet{K}(\lambda) ~~ \prod_{k=1}^\enne
\sigma_k(\lambda-\hbar(\enne-k))\;.
   \label{eval-sdet}
\end{eqnarray}

\subsection{An automorphism of the twisted Yangian}

The following map
\begin{eqnarray}
\label{check} \cS(\lambda)\longrightarrow   \Check
    \cS(\lambda)&=&\frac{\sdet \cS(-\lambda+\hbar(\enne/2-1-\rho))}
    {\sdet{K}(-\lambda+\hbar(\enne/2-1-\rho))} ~~\;
    \cS^{-1}(-\lambda-\hbar(\enne/2+\rho))\\
    &=&\qdet\cT(-\lambda+\hbar(\enne/2-1-\rho)) ~~
    \qdet\cT(\lambda+\hbar\enne/2)\nonumber\\
    &&\hspace{1cm}\times
    \big(\cT(\lambda+\hbar\enne/2)^t\big)^{-1}~K^{-1}~\cT^{-1}(-\lambda-\hbar(\enne/2+\rho))
\end{eqnarray}
is an automorphism of the twisted Yangien \cite{twmolev}. In order
to deal with analytical entries, we change the normalization of
$\Check S(\lambda)$ as follows:
\begin{eqnarray}
\widetilde
\cS(\lambda)=\prod_{k=1}^{\enne-1}\left(
\prod_{i=1}^\ell\Big(\lambda+a_i-\hbar(\enne/2-k)\Big)
\Big(-\lambda+a_i-\hbar(\enne/2-k+\rho)\Big)\right)~\Check \cS(\lambda)
\end{eqnarray}
\\
One can show that the vector $v^+$ is also a highest weight vector
for $\wt S(\lambda)$ with
\begin{eqnarray}
    && \wt S_{jk}(\lambda)\;v^+=0 \,,\qmbox{for} 1\leq k<j \leq \enne
    \label{HWmonotran11} \\
    && \wt S_{kk}(\lambda)\;v^+= \frac{1}{\zeta_k}
    ~~\wt \sigma_k(\lambda)\,v^{+}
    \label{HWmonotran21} \,, \qmbox{for}  1\leq k \leq \frac{\enne+1}{2}
    \qquad
\end{eqnarray}
and for $\frac{\enne}{2}+1\leq k \leq \enne$,
\begin{eqnarray}
    \label{HWmonotran31} &&\wt S_{kk}(\lambda)\;v^+=
    \frac{\theta\hbar}{2\lambda+\hbar\rho}\wt S_{\bar k \bar
    k}(\lambda)\,v^{+}+\frac{2\eps\lambda+\hbar(\eps\rho-\theta)}
    {2\lambda+\hbar\rho}\wt S_{\bar k \bar
    k}(-\lambda-\hbar\rho)\,v^{+}\,.\qquad
\end{eqnarray}
where
\begin{eqnarray}
\label{sig-tilde}
    \wt \sigma_k(\lambda) =
    \prod_{j=1}^{k-1}\sigma_j(-\lambda-\hbar(\enne/2+\rho-j))
    \prod_{j=k+1}^{\enne}\sigma_j(-\lambda-\hbar(\enne/2+\rho-j+1))\,.
\end{eqnarray}
The equality (\ref{HWmonotran21}) is computed similarly to
(\ref{HWmonotran2}). The relation (\ref{HWmonotran31}) follows from
 the crossing relation (\ref{sym-rel}) satisfied by $\Check \cS(\lambda)$.
 Note that the functions $\wt\sigma_{k}(\lambda)$ also obey the
 crossing relation
\begin{equation}
        \wt\sigma_{\bar k}(-\lambda-\hbar\rho)=\wt\sigma_{k}(\lambda)\,.
        \label{crostSig}
\end{equation}

The transfer matrix $\wt s(\lambda)=tr_a \wt{\cS}_a(\lambda)$
satisfies the commutation relations
\begin{eqnarray}
    [\wt s(\lambda),\wt s(\mu)]=0 \qmbox{,} [\wt s(\lambda),\wh s(\mu)] = 0
\end{eqnarray}
and the crossing relation
\begin{eqnarray}
\label{cross-tilde} \wt s(\lambda)=
\frac{2\lambda\;\varepsilon+\hbar(\rho\;\varepsilon-\theta)}
{2\lambda+\hbar(\rho-\theta)} ~~\wt s(-\lambda-\hbar\rho)\;.
\end{eqnarray}
 The Liouville contraction \cite{MNO} allows
us to find a constraint between $\wh s(u)$ and $\wt s(u)$:
\begin{eqnarray}
    \label{fus1}
    \wt s(u)\wh s(u-\hbar\enne/2) =
    \left(1+\frac{\hbar\theta}{2u-\hbar(\enne-\rho)}\right)
    \left(\varepsilon-\frac{\hbar\theta}{2u+\hbar\rho}\right)
    \prod_{k=1}^\enne \sigma_{\overline{k}}(u+\hbar(\enne/2+k))+s_\ff(u)\;.
\end{eqnarray}
This relation is computed by using the following equality (see ref.
\cite{MNO})
\begin{eqnarray}
    && \frac{Q_{12}}{\enne}R_{12}(-2u+\hbar(\enne-\rho))\wt \cS_1(u)
    R_{12}(2u+\hbar\rho) \wh\cS_2(u-\hbar\enne/2)
    \nonumber\\
    && \hspace*{2cm} =\frac{Q_{12}}{\enne}
    \left(1+\frac{\hbar\theta}{2u-\hbar(\enne-\rho)}\right)
    \left(\varepsilon-\frac{\hbar\theta}{2u+\hbar\rho}\right)
    \prod_{k=1}^N\sigma_{\overline{k}}(u+\hbar(\enne/2+k))\qquad
\end{eqnarray}
where $Q_{12} = \cP_{12}^{t_{1}}$ and $Q_{12}/\enne$ is a one-dimensional
projector.

\subsection{Analytical Bethe ansatz}

\subsubsection{Pseudo-vacuum}

As in the case of the closed spin chains, the first step of the analytical
Bethe Ansatz consists in finding a particular eigenvalue of the transfer
matrix $s(\lambda)$. This eigenvalue is computed thanks to the highest
weight vector $v^+$. Indeed, one gets
\begin{eqnarray}
\wh s(\lambda)\;v^+=\sum_{k=1}^\enne \wh {S}_{kk}(\lambda)v^+
=\Lambda^0(\lambda)\;v^+
\end{eqnarray}
where
\begin{eqnarray}
\Lambda^0(\lambda)=\sum_{k=1}^\enne g_k(\lambda)~\sigma_k(\lambda)\;.
\end{eqnarray}
The functions $g_k(\lambda)$ depend on the boundary and are given by \begin{eqnarray}
g_k(\lambda)=
\begin{cases}
\displaystyle \zeta_k~
\frac{2\lambda+\hbar(\rho+\theta)}{2\lambda+\hbar\rho}
\qmbox{,}  1\leq k \leq \frac{\enne}{2} \;,\vspace{3mm}\\
\zeta_{k}\qmbox{if} \enne\mbox{ is odd and } k=\frac{\enne+1}{2}
\vspace{3mm}\\
\displaystyle \zeta_k~
\frac{2\lambda+\hbar(\rho-\theta\varepsilon)}{2\lambda+\hbar\rho}
\qmbox{,} \frac{\enne}{2}+1\leq k \leq \enne
\end{cases}
\end{eqnarray}
while the functions $\sigma_k(\lambda)$ depend on the choice of the
representation and are given by (\ref{sigma}). Note that, although
$g_{k}(\lambda)$ has a pole at $\lambda=-\hbar\rho/2$, the residue
 $res_{\lambda=-\hbar\rho/2}\,\Lambda^{0}(\lambda)$ vanishes, so that
the eigenvalue is indeed analytical.\\
The functions $g_k(\lambda)$ satisfies the following relation
\begin{eqnarray}
g_k(\lambda)=\frac{2\lambda\varepsilon+\hbar(\rho\varepsilon-\theta)}
{2\lambda+\hbar(\rho-\theta)}~~
g_{\overline{k}}(-\lambda-\hbar\rho)\;.
\label{crossg}
\end{eqnarray}

\subsubsection{Dressing functions and fusion procedure}

The analytical Bethe ansatz states that all the eigenvalues of $\wh s(\lambda)$
can be written as
\begin{eqnarray}
    \Lambda(\lambda)=\sum_{k=1}^\enne
    g_k(\lambda)~\sigma_k(\lambda)~D_k(\lambda)\;,
\end{eqnarray}
where the dressing functions $D_k(\lambda)$ are rational functions and need
to be determined. The crossing relations (\ref{cross}),
(\ref{crossig}) and (\ref{crossg}) imply that
\begin{eqnarray}
    \label{crossD} D_{k}(\lambda)=
    D_{\overline{k}}(-\lambda-\hbar\rho)\qmbox{for}1\leq k \leq \enne\;.
\end{eqnarray}
This latter relation is sufficient to prove that the residue of
$\Lambda(\lambda)$ at $\lambda=-\hbar\rho/2$ vanishes.\\
In order to further constrain the dressing functions, we use the
Sklyanin determinant to fuse $\enne$ auxiliary spaces according to
$[\enne]^{\otimes \enne}=[1]\oplus \dots$. We start with the decomposition
\begin{eqnarray}
    {\cK}^{+}(\lambda)\, \cS_{<a_{\enne}\ldots a_{1}>}(\lambda)=
    {\cK}^{+}(\lambda)\, \cS_{<a_{\enne}\ldots
    a_{1}>}(\lambda)\,A_{\enne}+ {\cK}^{+}(\lambda)\,
    \cS_{<a_{\enne}\ldots a_{1}>}(\lambda)\,(1-A_{\enne})
\end{eqnarray}
where $\cS_{<a_{\enne}\ldots a_{1}>}(\lambda)$ has been introduced
in (\ref{SfusansK}),
\begin{eqnarray}
    &&{\cK}^{+}(\lambda) = \prod_{2\leq k\leq \enne}^{\longrightarrow}
    \left(
    R_{a_{k}a_{1}}^{t_{a_{k}}}(\lambda_{k}+\lambda_{1}+\hbar\rho)\cdots
    R_{a_{k}a_{k-1}}^{t_{a_{k}}}(\lambda_{k}+\lambda_{k-1}+\hbar\rho)
    \right)
\end{eqnarray}
and $\lambda_{k}=\lambda-\hbar(k-1)$.
Using the property (\ref{sdetSNP}) and taking the trace over the
auxiliary spaces $a_{1}$,\ldots, $a_{\enne}$, we get after some
calculation\footnote{The introduction of $\cK_{+}(\lambda)$ in the
process is essential to show that the l.h.s. is a function of the
transfer matrix solely.}
\begin{eqnarray}
    && \frac{2\lambda+\hbar(\rho-2\enne+2)}{2\lambda+\hbar(\rho-\enne+1)}\,
    \prod^{\enne-1}_{j=1}
    \frac{2\lambda+\hbar(\rho-2j+2)}{2\lambda+\hbar(\rho-2j+1)}\,
    \wh s(\lambda_{\enne})\cdots \wh s(\lambda_{1}) \nonumber\\
    && \hspace{6cm}=
    \frac{2\lambda+\hbar\Big(\rho-\left(3-\theta\right)n+1\Big)}
    {2\lambda+\hbar(\rho-2n+1)} \,\sdet\wh \cS(\lambda)
    +s_{\ff}(\lambda)\qquad
    \label{eq:somecalculation}
\end{eqnarray}
where $\displaystyle n=\left[\frac{\enne}{2}\right]$. Applying
relation (\ref{eq:somecalculation}) on an $\wh s(\lambda)$
eigenvector, we obtain using relation (\ref{sdet-qdet}) the fusion
relation for the soliton non-preserving dressing functions:
\begin{eqnarray}
    \label{sklyD} D_1(\lambda-\hbar\enne+\hbar)\dots D_\enne(\lambda)=1\;.
\end{eqnarray}

\subsubsection{Dressing functions for $\wt s(\lambda)$}

The entries of $\wt S(\lambda)$ are analytical. Therefore, we can
use the analytical Bethe ansatz to compute the eigenvalues of $\wt
s$. One gets
\begin{eqnarray}
\wt s(\lambda)\;v^+=\sum_{k=1}^\enne \wt {S}_{kk}(\lambda)v^+ =\wt
\Lambda^0(\lambda)\;v^+
\end{eqnarray}
where
\begin{eqnarray}
    \wt\Lambda^0(\lambda)=\sum_{k=1}^\enne \wt
    g_k(\lambda)~\wt\sigma_k(\lambda)\;.
\end{eqnarray}
The functions $\wt g_k(\lambda)$ depend on the boundary and are
given by
\begin{eqnarray}
    \wt g_k(\lambda)=\frac{1}{\zeta_k^2}~g_k(\lambda) \qmbox{for}
    1\leq k \leq \enne
\end{eqnarray}
and the functions $\wt \sigma_k(\lambda)$ are given by relation
(\ref{sig-tilde}). Let us remark, \textit{en passant}, that we have
the relations
\begin{eqnarray}
    \frac{\wt\sigma_{k+1}(\lambda)}{\wt\sigma_{k}(\lambda)}=
    \frac{\sigma_{k}(-\lambda-\hbar(\enne/2+\rho-k))}
    {\sigma_{k+1}(-\lambda-\hbar(\enne/2+\rho-k))}
\end{eqnarray}
which allows us to relate the
irreducibility criterion for
$\wt S(\lambda)$ to the ones of $S(\lambda)$, in accordance
with the automorphism (\ref{check}).

At this point, using the same ansatz than in the case of
$\Lambda(\lambda)$, the above eigenvalue can be dressed to obtain all the
eigenvalues of $\wt S(\lambda)$
\begin{eqnarray}
    \wt \Lambda(\lambda)=\sum_{k=1}^\enne \wt
    g_k(\lambda)~\wt\sigma_k(\lambda)\wt D_k(\lambda)\;.
\end{eqnarray}
Using the crossing relations (\ref{cross-tilde}), (\ref{crossg}) and
(\ref{crostSig}), the dressing
functions satisfy the following crossing:
\begin{eqnarray}
    \label{crosstD}
    \wt D_k(\lambda)=\wt D_{\overline{k}}(-\lambda-\hbar\rho)\;.
\end{eqnarray}
Through a fusion procedure with the Sklyanin determinant of $\wt S$, the
following constraint is obtained:
\begin{eqnarray}
    \label{sklytd}
    \wt D_1(\lambda-\enne\hbar+\hbar)\dots \wt
D_\enne(\lambda)=1\;.
\end{eqnarray}
Finally, from (\ref{fus1}) one gets
\begin{eqnarray}
    \wt D_\enne(\lambda) D_1(\lambda-\hbar\enne/2)=1\;. \label{DtD}
\end{eqnarray}

\subsubsection{Constraints for the dressing functions}

Guided by the results given in \cite{doikou1,aacdfr} for SNP spin chains in the
fundamental representation, we assume that the dressing functions are rational functions of the
form
\begin{eqnarray}
\label{dress:functD} D_k(\lambda)=\prod_{j=1}^{M^{(k-1)}}
\frac{\lambda+u_j^{(k-1)}}{\lambda-\lambda_j^{(k-1)}-\frac{\hbar\;(k-1)}{2}}
\frac{\lambda+v_j^{(k-1)}}{\lambda+\lambda_j^{(k-1)}-\frac{\hbar\;(k-1)}{2}}
\ \prod_{j=1}^{M^{(k)}}
\frac{\lambda+w_j^{(k)}}{\lambda-\lambda_j^{(k)}-\frac{\hbar\;k}{2}}
\frac{\lambda+x_j^{(k)}}{\lambda+\lambda_j^{(k)}-\frac{\hbar\;k}{2}}\;,
\end{eqnarray}
where $M^{(0)}=M^{(\enne)}=0$ and $u_j^{(k)}$, $v_j^{(k)}$,
$w_j^{(k)}$ and $x_j^{(k)}$ are coefficients to be determined. A
similar assumption is made for the form of $\wt D_k(\lambda)$. \\
Let us remark that the above form is also dictated by the SNP reflection
equation which shows that the exchange relations for the twisted
Yangian generators always imply both $R(\lambda-\mu)$ and
$R(\lambda+\mu)$. Having in mind the algebraic Bethe ansatz construction for the transfer matrix eigenvectors, we may conclude that the corresponding eigenvalues consist of terms of the form
$f(\lambda-\lambda_{j})\,f(\lambda+\lambda_{j})$, hence the form
(\ref{dress:functD}) assumed here.
\\
Imposing the constraints (\ref{crossD}) and (\ref{sklyD}) for the
dressing functions $D(\lambda)$, the constraints (\ref{crosstD}) and
(\ref{sklytd}) for the dressing functions $\wt D(\lambda)$ and
finally the constraint (\ref{DtD}) between the two types of dressing
functions, we obtain sufficient constraints to determine
$u_j^{(k)}$, $v_j^{(k)}$, $w_j^{(k)}$ and $x_j^{(k)}$ (and also the
$\wt D(\lambda)$ parameters) in terms of
$\lambda_j^{(k)}$. The dressing functions become,
 for $1\leq k\leq[\frac{\enne-1}{2}]$,
\begin{eqnarray}
\label{dress:functdDc} D_k(\lambda)=\prod_{j=1}^{M^{(k-1)}}
\frac{\lambda-\lambda_j^{(k-1)}-\frac{\hbar\;(k+1)}{2}}
{\lambda-\lambda_j^{(k-1)}-\frac{\hbar\;(k-1)}{2}}
\frac{\lambda+\lambda_j^{(k-1)}-\frac{\hbar\;(k+1)}{2}}
{\lambda+\lambda_j^{(k-1)}-\frac{\hbar\;(k-1)}{2}}
\prod_{j=1}^{M^{(k)}}
\frac{\lambda-\lambda_j^{(k)}-\frac{\hbar\;(k-2)}{2}}
{\lambda-\lambda_j^{(k)}-\frac{\hbar\;k}{2}}
\frac{\lambda+\lambda_j^{(k)}-\frac{\hbar\;(k-2)}{2}}
{\lambda+\lambda_j^{(k)}-\frac{\hbar\;k}{2}} \;.\quad
\end{eqnarray}
When $\enne=2n$ and $\rho\neq -\enne/2$, we have the following particular form for
$D_{n}(\lambda)$
\begin{eqnarray}
\label{dress:functDcn1} D_{n}(\lambda)&=&\prod_{j=1}^{M^{(n-1)}}
\frac{\lambda-\lambda_j^{(n-1)}-\frac{\hbar\;(n+1)}{2}}
{\lambda-\lambda_j^{(n-1)}-\frac{\hbar\;(n-1)}{2}}\
\frac{\lambda+\lambda_j^{(n-1)}-\frac{\hbar\;(n+1)}{2}}
{\lambda+\lambda_j^{(n-1)}-\frac{\hbar\;(n-1)}{2}}\nonumber\qquad\\
&&\hspace*{-2cm}\times\prod_{j=1}^{M^{(n)}}
\frac{\lambda-\lambda_j^{(n)}-\frac{\hbar\;(n-2)}{2}}
{\lambda-\lambda_j^{(n)}-\frac{\hbar\;n}{2}}\
\frac{\lambda+\lambda_j^{(n)}-\frac{\hbar\;(n-2)}{2}}
{\lambda+\lambda_j^{(n)}-\frac{\hbar\;n}{2}}\
\frac{\lambda-\lambda_j^{(n)}+\frac{\hbar\;(n+2+2\rho)}{2}}
{\lambda-\lambda_j^{(n)}+\frac{\hbar\;(n+2\rho)}{2}}
\frac{\lambda+\lambda_j^{(n)}+\frac{\hbar\;(n+2+2\rho)}{2}}
{\lambda+\lambda_j^{(n)}+\frac{\hbar\;(n+2\rho)}{2}}\qquad
\end{eqnarray}
whereas for $\enne=2n+1$, we have the following particular form for
$D_{n+1}(\lambda)$
\begin{eqnarray}
\label{dress:functdDcn1} D_{n+1}(\lambda)=\prod_{j=1}^{M^{(n)}}
\frac{\lambda-\lambda_j^{(n)}-\frac{\hbar\;(n+2)}{2}}
{\lambda-\lambda_j^{(n)}-\frac{\hbar\;n}{2}}\
\frac{\lambda+\lambda_j^{(n)}-\frac{\hbar\;(n+2)}{2}}
{\lambda+\lambda_j^{(n)}-\frac{\hbar\;n}{2}}\
\frac{\lambda-\lambda_j^{(n)}+\frac{\hbar\;(n+2+2\rho)}{2}}
{\lambda-\lambda_j^{(n)}+\frac{\hbar\;(n+2\rho)}{2}}
\frac{\lambda+\lambda_j^{(n)}+\frac{\hbar\;(n+2+2\rho)}{2}}
{\lambda+\lambda_j^{(n)}+\frac{\hbar\;(n+2\rho)}{2}}\quad
\end{eqnarray}
The other dressing functions, $D_k(\lambda)$ for
$[\frac{\enne+1}{2}]< k\leq \enne$, are determined by the crossing
relations (\ref{crossD}) and (\ref{crosstD}). \\
Note that when $\enne=2n$ and $\rho=-\enne/2$, one has to modify the
form (\ref{dress:functDcn1}). Indeed, in that case, the second line of 
(\ref{dress:functDcn1}) is a square, which has to be omitted. This 
leads to the simpler form:
\begin{eqnarray}
\label{dress:functDcnn1}&&\hspace{-1cm}
D_{n}(\lambda)=\prod_{j=1}^{M^{(n-1)}}
\frac{\lambda-\lambda_j^{(n-1)}-\frac{\hbar\;(n+1)}{2}}
{\lambda-\lambda_j^{(n-1)}-\frac{\hbar\;(n-1)}{2}}\
\frac{\lambda+\lambda_j^{(n-1)}-\frac{\hbar\;(n+1)}{2}}
{\lambda+\lambda_j^{(n-1)}-\frac{\hbar\;(n-1)}{2}}
\prod_{j=1}^{M^{(n)}}
\frac{\lambda-\lambda_j^{(n)}-\frac{\hbar\;(n-2)}{2}}
{\lambda-\lambda_j^{(n)}-\frac{\hbar\;n}{2}}\
\frac{\lambda+\lambda_j^{(n)}-\frac{\hbar\;(n-2)}{2}}
{\lambda+\lambda_j^{(n)}-\frac{\hbar\;n}{2}}\nonumber\\
\end{eqnarray}

\subsubsection{Bethe equations}

By imposing that the vanishing of the
$\lambda=\lambda_p^{(k)}+\frac{\hbar k}{2}$ residue of
$\Lambda(\lambda)$, we obtain a supplementary constraint between
the parameters $\lambda_j^{(n)}$ called Bethe equations. For
$1\leq k < \left[\frac{\enne+1}{2}\right]$, these equations read
\begin{eqnarray}
\frac{g_k(\lambda_p^{(k)}+\frac{\hbar k}{2})
\sigma_k(\lambda_p^{(k)}+\frac{\hbar k}{2})}
{g_{k+1}(\lambda_p^{(k)}+\frac{\hbar k}{2})
\sigma_{k+1}(\lambda_p^{(k)}+\frac{\hbar k}{2})} =-
\prod_{j=1}^{M^{(k-1)}}
\widehat{e}_{-1}(\lambda_p^{(k)},\lambda_j^{(k-1)})
\prod_{j=1}^{M^{(k)}}
\widehat{e}_{2}(\lambda_p^{(k)},\lambda_j^{(k)})
\prod_{j=1}^{M^{(k+1)}}
\widehat{e}_{-1}(\lambda_p^{(k)},\lambda_j^{(k+1)})\nonumber\\
\end{eqnarray}
where
\begin{equation}
    \widehat{e}_{x}(\lambda,\mu) =
    \frac{\lambda-\mu-\frac{\hbar x}{2}}{\lambda-\mu+\frac{\hbar x}{2}} \;
\frac{\lambda+\mu-\frac{\hbar x}{2}}{\lambda+\mu+\frac{\hbar x}{2}}
\end{equation}
The last equation ($\lambda=\lambda_p^{(n)}+\frac{\hbar n}{2}$
residue) depends on the parity of $\enne$ and on the choice of
$\rho$. \\
For $\enne=2n$ and $\rho\neq-\enne/2$, this equation reads
\begin{eqnarray}
&&\frac{g_n(\lambda_p^{(n)}+\frac{\hbar n}{2})
\sigma_n(\lambda_p^{(n)}+\frac{\hbar n}{2})}
{g_{n+1}(\lambda_p^{(n)}+\frac{\hbar n}{2})
\sigma_{n+1}(\lambda_p^{(n)}+\frac{\hbar n}{2})} =\nonumber\\
&&\hspace{-2cm}-\prod_{j=1}^{M^{(n-1)}}
\widehat{e}_{-1}(\lambda_p^{(n)},\lambda_j^{(n-1)})
\widehat{e}_{-1}(\lambda_p^{(n)}+\hbar(n+\rho),\lambda_j^{(n-1)})
\prod_{j=1}^{M^{(n)}}
\widehat{e}_{2}(\lambda_p^{(n)},\lambda_j^{(n)})
\widehat{e}_{2}(\lambda_p^{(n)}+\hbar(n+\rho),\lambda_j^{(n)})
\end{eqnarray}
In the particular case where $\rho=-\frac{\enne}{2}$, this equation
reduces to
\begin{eqnarray}
\frac{g_n(\lambda_p^{(n)}+\frac{\hbar n}{2})
\sigma_n(\lambda_p^{(n)}+\frac{\hbar n}{2})}
{g_{n+1}(\lambda_p^{(n)}+\frac{\hbar n}{2})
\sigma_{n+1}(\lambda_p^{(n)}+\frac{\hbar n}{2})} = -
\prod_{j=1}^{M^{(n-1)}}
\widehat{e}_{-1}(\lambda_p^{(n)},\lambda_j^{(n-1)})^2
\prod_{j=1}^{M^{(n)}}
\widehat{e}_{2}(\lambda_p^{(n)},\lambda_j^{(n)})
\end{eqnarray}
While for $\enne=2n+1$, it writes
\begin{eqnarray}
\frac{g_n(\lambda_p^{(n)}+\frac{\hbar n}{2})
\sigma_n(\lambda_p^{(n)}+\frac{\hbar n}{2})}
{g_{n+1}(\lambda_p^{(n)}+\frac{\hbar n}{2})
\sigma_{n+1}(\lambda_p^{(n)}+\frac{\hbar n}{2})} &=&-
\prod_{j=1}^{M^{(n-1)}}
\widehat{e}_{-1}(\lambda_p^{(n)},\lambda_j^{(n-1)})
\nonumber\\
&&\times\prod_{j=1}^{M^{(n)}}
\widehat{e}_{2}(\lambda_p^{(n)},\lambda_j^{(n)})\,
\widehat{e}_{-1}(\lambda_p^{(n)}+\hbar\frac{\enne+2\rho}{2},\lambda_j^{(n)})
\end{eqnarray} 

\section{Discussion}
Having specified the spectrum and Bethe ansatz equations for rational periodic 
and open spin chains in \cite{byebye} and in this article, the next natural step is 
the derivation of the spectrum and the Bethe ansatz equations for the 
$q$-deformed case, for both periodic and open boundaries. This may be 
achieved by employing the analytical Bethe ansatz techniques, however the 
ultimate goal is the formulation of a generic algebraic Bethe ansatz method for 
deriving not only the spectrum and Bethe ansatz for any irreducible 
representation of e.g ${\cal Y}(gl_{{\cal N}}))$, $U_{q}(gl_{{\cal N}})$, 
but also finding the corresponding eigenvectors. Such a process will exclusively
rely on the exchange relations emerging from the Yang--Baxter or reflection 
equation, depending on the choice of the boundary conditions. 
This project is of great mathematical and physical relevance, and is
under current investigation.

\vskip0.5cm

\textbf{Acknowledgements:} This work is supported by the TMR Network
`EUCLID. Integrable models and applications: from strings to condensed
matter', contract number HPRN-CT-2002-00325.

%\section{biblio}

\end{document}